\documentclass[aps,prb,twocolumn]{revtex4}
\usepackage{amssymb}
\usepackage{amsmath}
\usepackage{graphicx}

\begin{document}

\title{Vortex lattice and vortex bound states in CsFe$_2$As$_2$ investigated by scanning tunneling microscopy/spectroscopy}

\author{Xiong Yang, Zengyi Du, Hai Lin, Delong Fang, Huan Yang$^{*}$, Xiyu Zhu, and Hai-Hu Wen$^{\dag}$}

\affiliation{National Laboratory of Solid State Microstructures and Department of Physics, Collaborative Innovation Center of Advanced Microstructures, Nanjing University, Nanjing 210093, China}

\begin{abstract}
We investigate the vortex lattice and vortex bound states in CsFe$_2$As$_2$ single crystals by scanning tunneling microscopy/spectroscopy (STM/STS) under various magnetic fields. A possible structural transition or crossover of vortex lattice is observed with the increase of magnetic field, i.e., the vortex lattice changes from a distorted hexagonal lattice to a distorted tetragonal one at the magnetic field near 0.5 T. It is found that a mixture of stripelike hexagonal and square vortex lattices emerges in the crossover region. The vortex bound state is also observed in the vortex center. The tunneling spectra crossing a vortex show that the bound-state peak position holds near zero bias with STM tip moving away from the vortex core center. The Fermi energy estimated from the vortex bound state energy is very small. Our investigations provide experimental information to both the vortex lattice and the vortex bound states in this iron-based superconductor.
\end{abstract}

\maketitle

\section{INTRODUCTION}

The vortex state is a typical character of type-II superconductors. A vortex core carries one flux quantum $\Phi_0$, and is circulated by supercurrents consisting of Cooper pairs. The repulsive interaction between vortices will make them an ordered triangular lattice which is called an Abrikosov lattice in the sample with weak pinning. The vortex lattice can be easily altered by crystal defects or impurities which usually act as pinning centers\cite{Blatter1994}. In contrast, square vortex lattices have also been observed in some superconductors with fourfold symmetric lattice structure\cite{RNi2B2C1,RNi2B2C2,RNi2B2C3,V3Si,LSCO}. Vortex physics is very important for their relevance to the technological application of superconductors, and also for their relation to the superconducting mechanism of the material.

The low energy excitation within vortex cores has been predicted theoretically\cite{Caroli1964}, and the excitation bound states in a vortex core locate at different energies of $E_{\mu}=\pm\mu\Delta^{2}/E_F (\mu=1/2,\ 3/2,\ 5/2,\ ...)$, with $\Delta$ the superconducting gap and $E_F$ the Fermi energy. In most conventional superconductors, $E_F$ is many orders of magnitude larger than $\Delta$, causing a very small ratio of $\Delta/E_F$. Therefore, the energy difference $\Delta^{2}/E_F$ between neighboring energy levels is very small, thus the vortex bound states will have a continuous spatial evolution: A single peak at zero bias can be observed in the vortex center, and followed by a continuous separation to two peaks whose energies move gradually closer to $\pm\Delta$. This kind of vortex bound state was first observed\cite{Hess1990} in 2$H$-NbSe$_2$. In addition, the discrete vortex bound states were predicted theoretically\cite{QuantumLimit} in the system with small $E_F$, where the quantum limit condition is satisfied ($T/T_c\ll\Delta/E_F$). Such phenomenon was claimed to be observed \cite{YNiNC} in YNi$_2$B$_2$C.

The discovery of iron-based superconductors has motivated the worldwide investigations on their superconducting mechanism, which is very interesting because of the multiple electron and hole Fermi pockets\cite{DuMH2008} and possible sign-reversal $s\pm$ pairing\cite{Mazin2008,Kuroki2008,HuJP2008,Chubukov2009}. It is also worthy to investigate vortex physics in iron-based superconductors. The vortex dynamics has been widely studied previously\cite{vanderBeek,Demirdi2011,YangH2008,ShenB,GurevichReview,PuttiReivew}. The ordered hexagonal vortex lattice has been observed on KFe$_2$As$_2$\cite{KFeAs}, BaFe$_2$(As$_{0.67}$P$_{0.33}$)$_2$ \cite{BaFeAsP}, and optimally doped Ba$_{1-x}$K$_x$Fe$_2$As$_2$\cite{ShanL2011,YangH2012,HoffmanReview2011} at strong magnetic fields, while the order can be easily destroyed in some other materials\cite{Hoffman2009,Demirdi2011,Hanaguri2012,SongCL2013,DavisFeSeTe,MYChenFeSeTe}. Vortex bound states have also been detected in some iron-based superconductors\cite{Hoffman2009,ShanL2011,Hanaguri2012} with asymmetric peaks near the zero bias measured in the vortex center. The discrete vortex bound states of higher orders in the quantum limit are observed recently in FeTe$_{0.55}$Se$_{0.45}$ from our latest work\cite{MYChenFeSeTe}.

Comparing with some iron pnictides with both hole and electron Fermi pockets\cite{LiFeAs,NaFeCoAs,BaKFeAs}, AFe$_2$As$_2$ (A = K, Cs, or Rb) is in the extremely hole-doping level of 122 family and with only hole pockets. Meanwhile, it is assumed that the correlation effect in AFe$_2$As$_2$ is much stronger than the parent phase BaFe$_2$As$_2$. Therefore, the pairing symmetry of these materials may be different from the early proposed $s\pm$ pairing manner which requires nesting between the hole and electron pockets with almost equal sizes. Studies with different methods have suggested that there is a nodal superconducting gap in AFe$_2$As$_2$\cite{Hashimoto,XHChen,Okazaki,Taillefer,LiSY2013,LiSY2015}. In KFe$_2$As$_2$, a Van Hove singularity (VHS) appearing just a few meV below the Fermi energy is suggested to have an essential influence on the superconductivity\cite{FangDL2015}. All of these unusual features inspire us to investigate the rich and complex physics including the vortex physics in this family of heavily hole-doped materials.

Scanning tunneling microscopy/spectroscopy (STM/STS) is a useful technique for imaging the vortex core and studying the vortex bound states. In this paper, we present investigations on the vortices in CsFe$_2$As$_2$ single crystals by using STS/STM. We observed a structural transition or crossover of the vortex lattice with an increase of magnetic field. The ordered vortex lattice suggests the weak pinning in the material. We also observed the vortex bound states and try to obtain the coherence length from the tunneling spectra crossing a vortex.

\section{EXPERIMENTAL METHOD}

The single crystals of CsFe$_2$As$_2$ used in the experiments are synthesized by self-flux method\cite{Kikou2010}. The STM/STS are measured by a scanning tunneling microscope (USM-1300, Unisoku Co., Ltd.) with ultrahigh vacuum, low-temperature and high-magnetic field. The samples were cleaved at room temperature in an ultrahigh vacuum, and then quickly transferred into the microscope head. During all STM/STS measurements, we use the electrochemically etched tungsten tips. To lower down the noise of the differential conductance spectra, a lock-in technique was used with an ac modulation of 0.8 mV at 987.5 Hz. The offset of the tunneling current is calibrated in the tip-withdraw status before the tunneling spectrum measurements. The offset bias for a set of tunneling spectra was calibrated by the averaged $I$-$V$ curves which are recorded under the same tip conditions, while the offset voltage is determined by the voltage value in the zero tunneling current, i.e., $V_\mathrm{offset}=V(I=0)$.

\section{RESULTS AND DISCUSSION}

\subsection{Tunneling spectra measured on CsFe$_2$As$_2$}

\begin{figure}
\includegraphics[width=8cm]{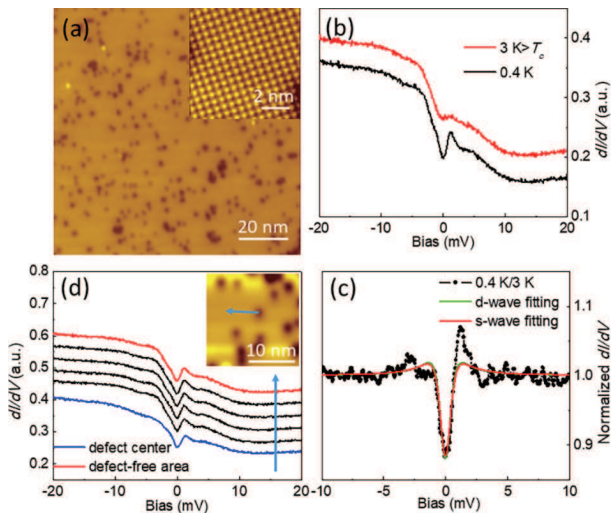}
\caption {(color online) (a) A typical topographic image of the Cs-terminated surface on CsFe$_2$As$_2$ measured with bias voltage of $V_{bias}=170$ mV and tunneling current of $I_t=50$ pA. The inset shows the atomically resolved image ($V_{bias}=20$ mV and $I_t=200$ pA). (b) Typical tunneling spectra measured at some defect-free position on CsFe$_2$As$_2$ at different temperatures. The spectra are offset for clarity. (c) Normalized tunneling spectrum measured at 0.4 K by the one measured at 3 K, and the Dynes model fitting curve by an $s$-wave and a $d$-wave gap. (d) A series of tunneling spectra across a defect signatured by a dark spot. The spectra are offset for clarity.
}\label{fig1}
\end{figure}

Figure~\ref{fig1}(a) shows a typical topographic image of the cleaved surface on a CsFe$_2$As$_2$ single crystal. Although various natural defects exist on the terminal surface, most of the areas are clean enough to image the square atomic lattice. The lattice constant is about 5.46 \AA, which is obtained from the atomically resolved image as shown in the inset of Fig.~\ref{fig1}(a). Here the lattice constant is consistent with $\sqrt{2}$ times of Cs lattice constant in the bulk sample, while the similar situation has been observed in KFe$_2$As$_2$ \cite{FangDL2015}. This phenomenon may be explained by the reconstruction of half remaining Cs atoms in one layer to maintain the  charge-neutral surface after cleavage\cite{HoffmanReview2011}. From the density and configuration of the alkali metal atoms on the surface, we assume a self-consistent model\cite{FangDL2015} to interpret the surface morphology, namely the main axes of the surface reconstructed atoms are the same as that of Fe atoms in the layer beneath.

Figure~\ref{fig1}(b) shows temperature dependent tunneling spectra measured at some defect-free locations. In KFe$_2$As$_2$, a sharp peak was observed with peak position a few meV below the Fermi level, which is proved to be induced by the VHS\cite{FangDL2015}. However, the peak of VHS is not observed on any spectra measured on CsFe$_2$As$_2$. This suggests that, the VHS may be very far from the Fermi energy or even disappear due to the change of the band structures. It also suggests that the increasing ionic size of the alkali metal may have a greater effect on the electronic properties than previous theoretical estimation\cite{AFe2As2Band}. Nevertheless, the spectrum on CsFe$_2$As$_2$ shows an asymmetric background with more contribution from the occupied states, which is similar to the situation in KFe$_2$As$_2$ and consistent with theoretical prediction\cite{AFe2As2Band,Mizukami2016}. The bulk $T_c$ of CsFe$_2$As$_2$ is about 2.1 K, as judged from the zero-resistance temperature\cite{YangH2016}. However, the spectrum measured at 3 K still shows some dip feature at zero bias. This phenomenon may be induced by the possible pseudogap feature, which has been discussed in our previous work\cite{YangH2016}. The energy difference between the two coherence peaks is about 1.2 meV on the normalized spectrum shown in Fig.~\ref{fig1}(c). It is very difficult to determine the exact gap value and the gap structure from the tunneling spectra with such high zero-bias differential conductance. We thus tried to use the Dynes model\cite{Dynes} with an $s$-wave or a $d$-wave gap to fit the normalized spectrum, and obtained $\Delta=0.42$ meV for the $s$-wave gap and $\Delta_0=0.57$ meV for the $d$-wave gap. The corresponding values of $2\Delta/k_BT_c$ are 4.6 and 6.3, respectively, and both of the two values are larger than 3.53 predicted by the Bardeen-Cooper-Schrieffer (BCS) theory in the weak coupling regime. Since several parameters as well as different gap structures are involved in the fitting, there are certainly some error bars for the determined gap values. The gap values obtained here are slightly larger than those estimated from bulk specific heat measurements in the sample with a slightly lower $T_c$\cite{XHChen}. It should be noted that CsFe$_2$As$_2$ is a multiband superconductor, and a one-gap model is not enough to describe the gap feature. Therefore, further detailed investigation is obviously required to determine the exact gap structures in this material as in KFe$_2$As$_2$ by high-resolution angle-resolved photoemission spectroscopy (ARPES) measurement\cite{Okazaki}.

Although there are many defects shown by the dark spots on the topography, they seem not to affect the superconducting feature judged from the quite uniform tunneling spectra measured along an arrowed line crossing a defect in Fig.~\ref{fig1}(d). Therefore, one can conclude that the pinning strength to the vortices by these defects may be also very weak.

\subsection{Possible structural transition of vortex lattice}

\begin{figure}
\includegraphics[width=8cm]{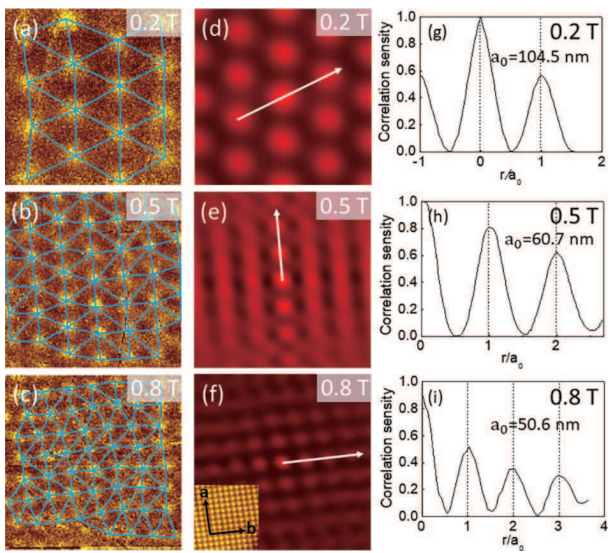}
\caption {(color online) (a)-(c) Images of vortex lattices measured at 0.4 K under various magnetic fields. The scanning areas are all 380 $\times$ 380 nm$^2$. Delaunay triangulation plots are used to indicate the lattice structure in the figures. (d)-(f) Self-correlation mapping to the real-space vortex lattice image measured at different fields in (a)-(c), respectively. The inset in (f) shows the atomically resolved surface which is a part of the vortex image region, and the main axes of the surface Cs atoms are marked by black arrows. (g)-(i) The spatial evolution of intensity along the arrowed lines in (d)-(f), respectively. The spatial period $a_0$ is also presented for each figure.
} \label{fig2}
\end{figure}

The high density of states within the vortex core or the vortex bound state make it possible to image the vortices by using zero-bias mapping of STM measurements. Figures~\ref{fig2}(a)-\ref{fig2}(c) show the vortex lattices measured at 0.2, 0.5 and 0.8 T, respectively. The vortices are all round shaped, and this may suggest the almost isotropic superconducting gap and Fermi velocity. The distances between neighboring vortices are roughly the same, which is more clear in the Delaunay triangulation plots in the figures. The nearly uniformly-spaced vortices denote that the pinning effect is very weak in the sample. From the vortex images in our experiment, one can see that the vortex lattice is hexagonal at 0.2 T, and then it changes to a square lattice with slight distortion at 0.8 T. Figures~\ref{fig2}(d)-\ref{fig2}(f) present the self-correlation patterns to the real-space images, and the structure transition or crossover can be clearly observed from the sixfold symmetric pattern to the distorted fourfold symmetric pattern. The vortex lattice constant $a_0$ can also been calculated from the self-correlation images, and the values are 104.5, 60.7, and 50.6 nm at 0.2, 0.5 and 0.8 T, respectively. The error bar is about 3.0 nm defined as the distance between two neighbored pixels in the figure. With no doubt, the vortex lattice constant is shrunk as expected with the increase of magnetic field. Concerning the average intervortex spacing, we did the calculation based on a square lattice and found $a_0=101.7$, 64.3 and 50.8 nm corresponding to the fields of 0.2, 0.5 and 0.8 T, respectively. These values are close to the experimental results.

\begin{figure}
\includegraphics[width=8cm]{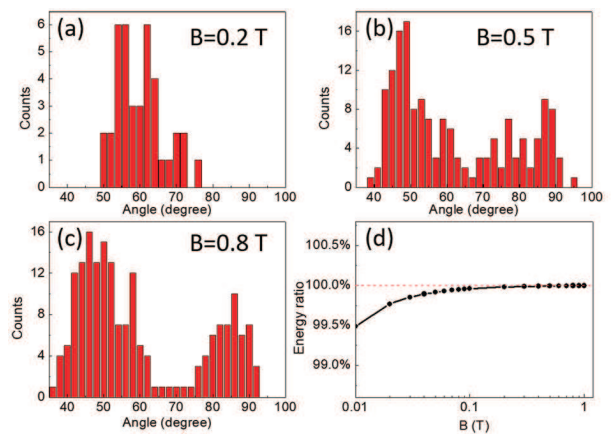}
\caption {(color online) (a)-(c) Statistics on the angle values of all the Delaunay triangles in Figs.~\ref{fig2}(a)-\ref{fig2}(c). (d) The ratio of vortex interaction energy per unit length in the hexagonal lattice over that in the tetragonal lattice (see text).
} \label{fig3}
\end{figure}

Furthermore, we do the statistics on the angle values of all the Delaunay triangles in Figs.~\ref{fig2}(a)-\ref{fig2}(c), and the results are shown in Figs.~\ref{fig3}(a)-\ref{fig3}(c). One can see that the statistics on the angle values versus magnetic field give a consistent expectation for a crossover from hexagonal to square lattice. At 0.2 T, the angle values are all concentrated near 60 degrees, which is the angle of the sixfold hexagonal lattice. When the magnetic field increases, the angle components near 45 and 90 degrees appear, and finally dominate at 0.8 T. From the points mentioned above, we conclude that there is a structural transition of the vortex lattice from a hexagonal to a distorted square lattice.

Actually the first order transition from hexagonal to square vortex lattice was investigated\cite{Suzuki} by using nonlocal London theory and Eilenberger theory, and it was argued as a generic phenomenon for a system with tetragonal structure. While since the iron based superconductors have the multiband feature, it is desired to know whether this conclusion is still totally or to some extent valid. In the following we use a simple simulation based on the interaction between vortices to argue that this transition is a generic feature. The repulsive interaction between two vortices can result in an interaction energy, and the energy for the $i^\mathrm{th}$ vortex can be expressed as \cite{Tinkham}
\begin{equation}
E^{i}=\sum_{j\neq i}\frac{\Phi_0^2}{8\pi^2\lambda^2}K_{0}\left(\frac{r_{ij}}{\lambda}\right).
\end{equation}
Here $K_{0}(x)$ is the lowest-order of modified Bessel function, and the penetration depth $\lambda_{ab}=0.3\ \mu$m from a previous report\cite{Mizukami2016} in CsFe$_2$As$_2$. The $r_{ij}$ stands for the distance between $i^\mathrm{th}$ and $j^\mathrm{th}$ vortex, and we do the calculation to all the vortices with $r_{ij}<100\lambda_{ab}$ for the hexagonal and square lattices, respectively. The field dependent ratio of the interaction energy in the hexagonal lattice to that in the square lattice is shown in Fig.~\ref{fig3}(d). One can clearly see that the interaction energy is lower for the hexagonal lattice than for the square lattice at low magnetic fields, therefore the lattice prefers the hexagonal one. When the magnetic field increases, the interaction energy difference between the two kinds of lattices becomes much smaller, and is negligible ($<$0.01\%) when the field is above 0.5 T. Qualitatively, from this point of view, it is not strange that the transition from a hexagonal to square lattice is observed in this sample. In addition, the square vortex lattices have been observed in some superconductors with fourfold symmetric structure\cite{RNi2B2C1,RNi2B2C2,RNi2B2C3,V3Si,LSCO}, hence it is normal that the same phenomenon has been observed on CsFe$_2$As$_2$ with tetragonal crystal structure. A proof of this issue is the consistence of the main axis directions of the vortex lattice and those of the surface Cs lattice, which is clearly displayed in Fig.~\ref{fig2}(f). It should be noted that the main axes of the surface reconstructed Cs atoms are the same as that of Fe atoms of the sample from our point of view\cite{FangDL2015}.

The Fermi surface shape\cite{Kogan} and the variations in the gap amplitude or gap structure\cite{Ichioka} will give influence on the form of the vortex structure. In this sample, the repulsive energy values are very close to each other for hexagonal and square lattices from the calculation as described above. The square lattice at a high magnetic field may be the only choice for a Fermi surface contour with fourfold symmetry. Recently, a detailed numerical simulation was carried out in type-II superconductors with a fourfold anisotropy, and the structural transition from a triangular lattice to a square one has been obtained with increasing of vortex-vortex interaction or the equivalent increasing of magnetic field\cite{NJPvortexcal}. This is consistent with our experimental results.

It should be mentioned that an isotropic hexagonal vortex lattice was observed in its sister compound KFe$_2$As$_2$ by neutron scattering, which seems to be different from the present situation in CsFe$_2$As$_2$. One reason for this difference may be from the different electronic structures, for example in KFe$_2$As$_2$, we observed a strong von Hove singularity near Fermi energy\cite{FangDL2015}, while it is absent in CsFe$_2$As$_2$\cite{YangH2016}. An alternative reason is the difference of superconducting gap structures in these two materials.

\subsection{Stripe-like mixture of hexagonal and square vortex lattice}

\begin{figure}
\includegraphics[width=8cm]{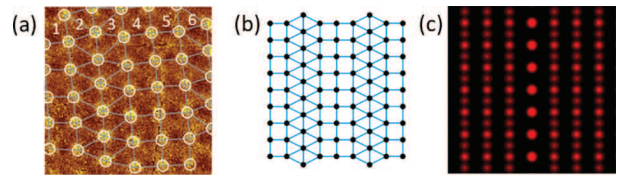}
\caption {(color online) (a) Experimental result of stripe-like vortex lattice mixed by hexagonal and square lattice. The data are the same as those in Fig.~\ref{fig2}(b), while the intervortex plots are a bit different to show the strip features. (b) The schematic image of the possible stripe-like vortex lattice structure in the crossover region. (c) Self-correlation image of (b).
} \label{fig4}
\end{figure}

We also notice that the vortex lattice seems to be a mixture of stripe-like hexagonal and square lattices at magnetic field of 0.5 T. The self-correlation map in Fig.~\ref{fig2}(e) also shows some stripe-like features along one of the crystallographic axes. Combining with the plots of interconnections between vortices in Fig.~\ref{fig4}(a), we can see six columns of vortices as marked by numbers in the image. Here, the 2$^\mathrm{nd}$, 3$^\mathrm{rd}$ and 6$^\mathrm{th}$ columns from left form stripes with distorted hexagonal structures, while the other three columns form stripes of square lattice. Since the hexagonal lattices are distorted in the stripes, the enclosed angles of vortices in this part are usually deviating from 60 degrees. This is why the statistic peak intensity is not strong at 60 degrees, but behaves as a rather wide distribution from 55 to 75 degrees, see Fig.~\ref{fig3}(b). According to the measured vortex lattice, we plot the possible schematic image of the artificial vortex lattice in Fig.~\ref{fig4}(b). Such artificial lattice is combined with hexagonal or square lattice stripes, and the corresponding self-correlation result shown in Fig.~\ref{fig4}(b) is very similar to the resultant data from experiment. Hence, we have observed a vortex structure with the mixture of strip-like tetragonal and hexagonal lattices, which is different from the random mixture of the triangular and square lattices and may be a new vortex lattice phase in the crossover region. The vortex structures have been studied by small-angle neutron-scattering studies previously in the crossover regions\cite{RNi2B2C3,V3Si}, and different domains with different kinds of ordered lattices are proved to coexist in the crossover regions. However, our direct observation of the strip-like combinations by two kinds of coexisted lattices provides a second possibility of the vortex structure in the crossover region. Here we must emphasize that, since the field of view in STM measurements is quite limited, we have no evidence for a long range order of such configuration. Recently, the similar combination of square and triangular ordering chains was predicted by the numerical simulation in a fourfold anisotropic superconductor, and such obtained construction is in the form of a kind of Archimedean tiling\cite{NJPvortexcal}. This can be regarded as the theoretical support to our experimental results and conclusions.

\subsection{Vortex bound state}

\begin{figure}
\includegraphics[width=8cm]{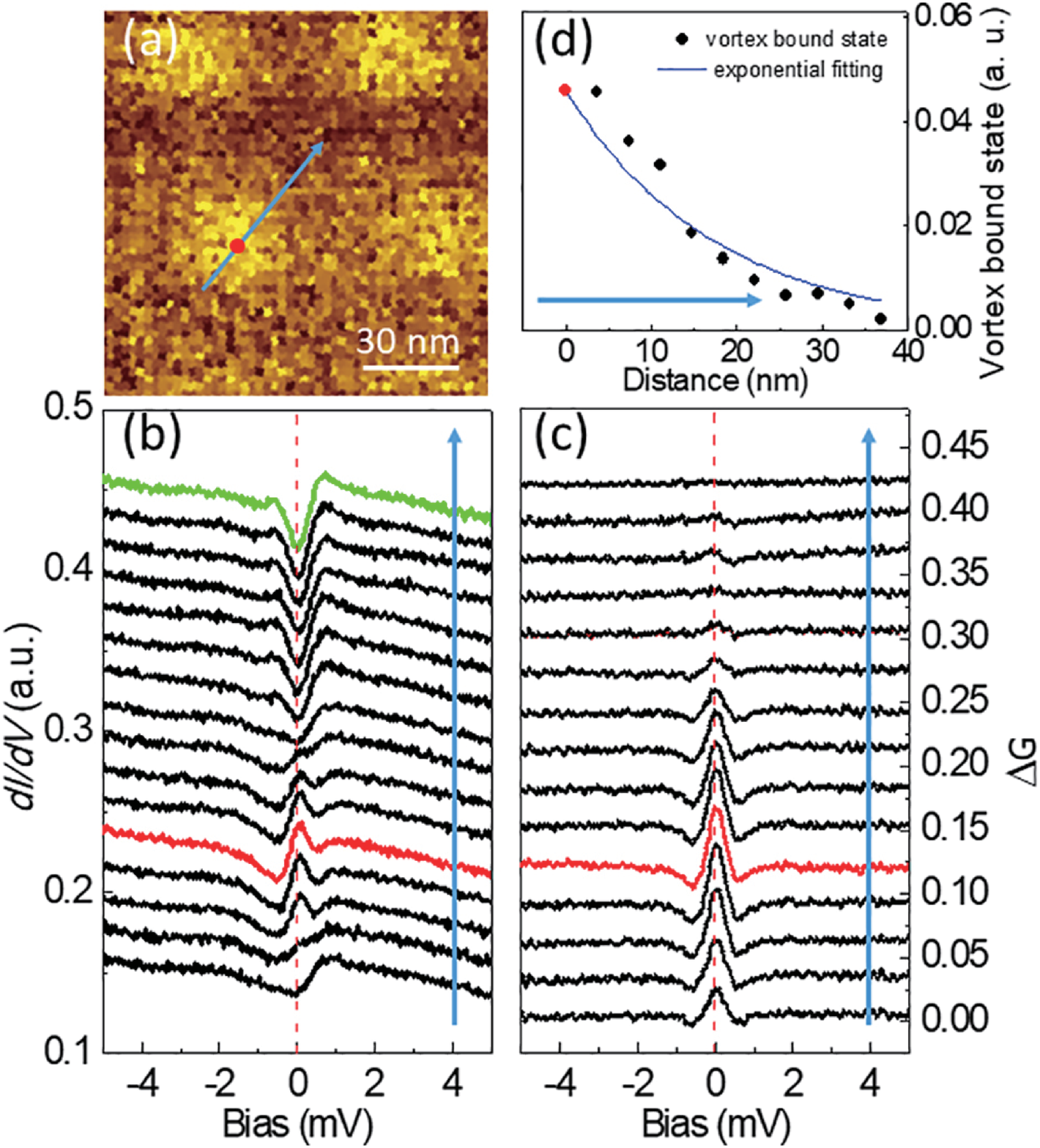}
\caption {(color online) (a) Vortex image by mapping of DOS at zero-bias. (b) Spatial evolution of the tunneling spectra at 0.4 K crossing a vortex core center along the arrowed line in (a). The spectra are offset for clarity. The red curve is measured at the vortex center, while the green one is measured far away with a distance of 40.5 nm from the vortex center. (c) The subtracted results $\Delta G$ of the tunneling spectra with the green curve in (b). The bound-state peak location is very close to zero bias (the red vertical dashed line). (d) Spatial evolution of $\Delta G$ taken at 0.05 mV from the difference spectra in (c) and fitting curve of the exponential decay formula.
} \label{fig5}
\end{figure}

In order to get a deeper insight into the vortex core state, we study the tunneling spectra crossing an individual vortex. Figure~\ref{fig5}(b) shows the spatial evolution of the tunneling spectra along the arrowed line in Fig.~\ref{fig5}(a), and one can observe the obvious vortex-bound-state peak on the spectrum measured exactly at the vortex center. The green curve in Fig.~\ref{fig5}(b) is the spectrum measured far away from the vortex center. Considering the asymmetric background of the spectra in CsFe$_2$As$_2$, we use this tunneling spectrum as the background and subtract all other curves with it. The difference $dI/dV$ spectra $\Delta G$ are presented in Fig.~\ref{fig5}(c), with a clear vortex bound-state peak locating at a rather small energy of about +0.05 mV on each curve. The peak energy almost doesn't shift when the STM tip moves away from the vortex center, which is similar to the situations in LiFeAs\cite{Hanaguri2012} and FeTe$_{0.55}$Se$_{0.45}$ \cite{MYChenFeSeTe}. The bound state peak observed here should be constructed by two peaks of positive and negative energies near zero-bias. Now the observed single peak locates at a slight positive energy, which may be induced by a stronger intensity of the peak at the positive energy than that of the negative one, if they would not merge together. In order to estimate the Fermi energy, we attribute the peak energy to the lowest level of bound states, i.e., $E_{1/2}=0.05$ meV. Considering the superconducting gap 0.42 meV for the single $s$-wave gap and 0.57 meV for the single $d$-wave gap, we can obtain $E_F= 1.7$ or 3.2 meV, respectively. This is really a small value, which indicates that CsFe$_2$As$_2$ may have very small Fermi energy. The value of $E_F$ derived here is about several meV, which is close to that estimated from the ARPES data for the four holelike $\varepsilon$ pockets near the Brillouin zone corners\cite{ARPESCsFe2As2}. However, the peak energy 0.05 meV is really near the energy resolution of our experiment. Thus, the estimation of $E_F$ may have some error.

Usually the coherence length can be estimated by the vortex core size from STM measurements. The way often adopted\cite{Hoffman2009} is through fitting the spatial evolution of zero-bias conductance by the exponential decay formula, where $r$ is the absolute value of distance from the vortex center. Since the bound state peak locates very close to zero-bias, we take the data of difference $dI/dV$ at 0.05 mV for coherence length analysis and show the data of $\Delta G(r)$ in Fig.~\ref{fig5}(d). Then we fix the $\Delta G_0=\Delta G(r=0)$ as a constant, and use the formula $\Delta G(r) = \Delta G_0\exp(-r/\xi_b)$ to fit the experimental data, and finally obtain the calculated value $\xi_b=17.3\pm1.8$ nm. Obviously, the value of $\xi_b$ is close to the value of 16.6 nm estimated from the upper critical field $H_{c2}=1.2$ T at 0.4 K\cite{LiSY2013}. It should be noted that the exponential-decay fitting method is an empirical protocol that provides only a rough estimate of the coherence length, although the estimated value here is close to the one obtained from the upper critical field\cite{LiSY2013}.

\section{SUMMARY}

To summarize, we perform STM/STS investigations on vortices in the iron-based superconductor CsFe$_2$As$_2$. We observe a structural transition or crossover of vortex lattice from hexagonal to square lattice near 0.5 T. This vortex lattice structural transition or crossover can be understood by considering the inter-vortex interaction and the crystal structure. The intermediate lattice structure is the combination of stripe like hexagonal and square lattices, which may be a new type of vortex lattice structure. In addition, we observe a non-splitting vortex bound state peak located near zero bias. Our observations of vortex physics in CsFe$_2$As$_2$ will help to understand the vortex dynamics and fundamental physics in iron-based superconductors.

\begin{acknowledgments}

This work was supported by National Key Research and Development Program of China (Grant No. 2016YFA0300401), National Natural Science Foundation of China (Grant No. 11534005), and Natural Science Foundation of Jiangsu (Grant No. BK20140015).

\end{acknowledgments}

$^*$ huanyang@nju.edu.cn

$^{\dag}$ hhwen@nju.edu.cn

\end{document}